# Are Trump and Bitcoin Good Partners?

*Jamal BOUOIYOUR* [†] *and Refk SELMI* [‡]

## Abstract

During times of extreme market turmoil, it is acknowledged that there is a tendency towards "flight to safety". A strong (weak) safe haven is defined as an asset that has a significant positive (negative) return in periods where another asset is in distress, while hedge has to be negatively correlated (uncorrelated) on average. The Bitcoin's surge alongside the aftermath of Trump's win in the 2016 U.S. presidential elections has strengthened its status as the modern safe haven. This paper uses a truly noise-assisted data analysis method, termed as Ensemble Empirical Mode Decomposition-based approach, to examine whether Bitcoin can act as a hedge and safe haven for U.S. stock price index. The results document that the Bitcoin's safe-haven property is time-varying and that it has primarily been a weak safe haven in the short term and the long-term. We also demonstrate that precious metals lost their safe haven properties over time as the correlation between gold/silver and U.S. stock price declines from short- to long-run horizons.

[†] CATT, University of Pau; *Corresponding Author:* Avenue du Doyen Poplawski, 64000 Pau, France; Tel: +33 (0) 5 59 40 80 01, Fax: +33 (0) 5 59 40 80 10, E-mail: jamal.bouoiyour@univ-pau.fr

[‡] University of Tunis; University of Pau; E-mail: s.refk@yahoo.fr

# 1. Introduction

Bitcoin was created in 2009 by an anonymous programmer under the pseudonym Satoshi Nakamoto and has since achieved a widest level of international recognition. Given the growing attention given to Bitcoin, researchers have revolved around its properties by asking several questions. One can cite: Is Bitcoin an alternative currency? Is it a speculative bubble? Is it a safe haven? Is it a hedge? Is it a portfolio diversifier? etc…

The majority of studies supported that Bitcoin is likely to be a speculative bubble rather than a safe haven or a hedge (Bucholz et al. 2012; Kristoufek 2013; Ciaian et al. 2014; Yermack 2014; Bouoiyour et al. 2015; Bouoiyour and Selmi 2015). Some researchers attributed this great speculation to the fact that there is no guarantee of repayment at any time (Kristoufek 2013; Yermack 2014). Bitcoin is also not yet accepted as a payment system across large markets and does not have an underlying value derived neither from consumption nor production process such as the precious metals including gold (Glaster et al. 2014). In addition, being a virtual currency, Bitcoin is highly sensitive to cyber-attacks, which can easily destabilise the Bitcoin system and then generate excessive price volatility (Ciaian et al. 2016). Another potential element that may explain the speculative attitude of Bitcoin is its strong dependence to media coverage. Accordingly, Lee (2014) showed that the alteration of positive and negative news contributed to high Bitcoin price cycles. Moreover, exponents of Bitcoin's virtues indicated that as it is an alternative currency and transaction tool that does not depend on a network of financial institutions, users are insulated against the untoward risks. This seems relevant when we consider that Bitcoin was created with the onset of the global financial collapse when governments and financial institutions lose public trust. This has give weight to compare Bitcoin to gold, as gold is perceived, in theory, as a hedge and safe haven to protect against similar risks. Baur and Lucey (2010) and Baur and McDermott (2010) claimed that there exists a significant linkage among gold price and financial assets. They defined a safe haven and a hedge depending on the sign and the significance of the coefficients associated to the gold price. The property of a safe-haven asset is to deliver or retain its value during times of crisis. So, the primary characteristic of a safe haven is the specific period in which the return is non-negative, whereas hedge has to be negatively correlated (or uncorrelated). In this context, Dyhberg (2015) tried to test the hedging capabilities of Bitcoin by addressing whether Bitcoin is a virtual gold. The study suggests that Bitcoin can serve as a hedge against stocks over shortest horizons. It concludes that Bitcoin possess hedging characteristics as gold and can be incorporated in a portfolio to mitigate the harmful effects of sudden shocks.

Even though the literature has largely documented how precious metals (with large extent gold) can act as a safe haven during financial turmoil (Baur and Lucey 2010; Bouoiyour and Selmi 2016), it ignored the role that may be played by Bitcoin as a hedging investment and a portfolio diversifier. Bitcoin is an alternative to mainstream currencies and the economy (Bouri et al. 2015). If traders and investors lose trust to currencies or to the whole economy, they may resort to Bitcoin. This is one of the potential causes behind the consideration of Bitcoin as a "digital gold" (Popper 2015). Despite its relevance, research examinaning Bitcoin as a diversifier, hedge, or safe heaven is extremely limited. To our best knowledge, there are only two studies that have attempted to answer this question (Bouri et al. 2015; Dyhberg 2015). This study tries to fill this gap by addressing whether Bitcoin join gold in hedging and safe haven status during the political uncertainty surrounding the 2016 U.S. presidential elections.

Financial markets had widely priced in a win for Clinton, who they viewed as a better short-run outcome because she represented few unknowns and thus less uncertainty. In response to the Trump's stunning triumph, the asset markets around the world plunged markedly as investors were concerned about evolving volatility. This has led to a trend towards questioning the effectiveness of standard economic and financial structures which govern the conventional monetary and financial system. Here, the digital currency (in particular, Bitcoin) is leading the charge by providing a completely decentralized secure alternative to fiat currencies during times of economic and geopolitical unrest. The price of Bitcoin starts 2017 as the top currency; it exceeds $ 1000 in First Day of 2017 trading for the first time on Coindesk Bitcoin Price Index since early January 2014.

Because the interaction of complex phenomena like Bitcoin and asset markets may mask the regularities we would like to identify, we apply a new data analysis tool, namely Ensemble Empirical Mode Decomposition (EEMD), which decomposes the Bitcoin price and the U.S. stock price index (in particular, S&P 500) into a scale-on-scale basis and at each scale it is estimated the correlation. The motivations behind the use of this technique arise in the desire to extract intrinsic characteristics inherent to the time series. In doing so, we show that hedge and safe-haven properties of Bitcoin and precious metals for U.S. stock market is time-varying. While Bitcoin serves as a weak safe haven (negative correlation) in shorter time scales, and as a hedge in the medium- and the long-run (insignificant correlation). We show also that gold and silver lost their hedge and safe haven properties as the correlation between precious metals and U.S. stock price is declining from one scale to another.

The outline of the paper is as follows: Section 2 describes our methodology. Section 3 reports the results. Section 4 discusses the findings and concludes.

## 2. Methodology and data

The traditional time series analysis tools usually rely on Fourier transforms in one way or another. Nevertheless, according to Huang et al. (1998), the Fourier transform might prompt inaccurate information owing essentially to the nature (in the time domain) of the transform. Wavelets have a problem of shift variance. Precisely, if the start point varies, by for example dropping the initial point, the wavelet transform may reveal distinct outcomes. However, the EMD method makes no assumption about linearity or stationarity and the intrinsic mode functions (IMFs) are often easily depicted[3]. A signal can be adaptively disentangled into a sum of finite number of zero mean oscillating components having symmetric envelopes defined by the local maxima and minima. The EMD is based on the sequential extraction of energy associated with distinct frequencies ranging from high fluctuating components (short-run) to low fluctuating modes (long-run). With the Hilbert transform, the IMF prompts instantaneous frequencies as functions of time that help to properly identify imbedded structures. The EMD aims at transforming the time series to hierarchical structure by means of the scaling transformations. In brief, it quantifies the changeability captured via the oscillation at different scales and locations.

In practice, the IMFs are decomposed by determining the maxima and minima of series $x(t)$, generating then its upper and lower envelopes ($e_{\min}(t)$ and $e_{\max}(t)$) with cubic spline interpolation. We first measure the mean $m(t)$ for different points from upper and lower envelopes:

$$m(t) = (e_{\min}(t) + e_{\max}(t))/2 \qquad (1)$$

We decompose the mean of the time series under study to identify the difference $d(t)$ between $x(t)$ and $m(t)$:

$$d(t) = m(t) - x(t) \qquad (2)$$

We present $d(t)$ as the $i^{th}$ IMF and we replace $x(t)$ with the residual $r(t) = x(t) - d(t)$. If not, we replace $x(t)$ with $d(t)$.

Then, we connect the local maxima with the upper envelope and the minima with the lower one. When residue meets the conditions that the number of zero-crossings and extrema do not differ by more than one and when the residue ($r$) becomes a monotonic function and data cannot be extracted into further intrinsic mode functions (Huang et al. 2003), the original time series can be denoted as the sum of some IMFs and a residue:

$$X(t) = \sum_{j=1}^{N} c_j(t) + r(t) \qquad (3)$$

---
[3] For detailed discussion of the EMD technique and comparison to other time series analysis tools, you can refer to Huang et al. (1998) and Flandrin et al. (2004).

The EMD technique (Huang et al., 1998) is illustrated more simply in Figure 1. The decomposition of the signal into IMFs is carried out as follows: After determining the positive peaks (maxima) and negative peaks (minima) of the original signal, we construct the lower and the upper envelopes of the signal by the cubic spline method (red). In addition, we measure the mean values (blue) by averaging the upper envelope and the lower envelope. Besides, we subtract the mean from the original signal to find the first intrinsic mode function (IMF1). Then, we calculate the first residual component by subtracting IMF1 component from the original signal. Finally, we repeat the steps above until the final residual component becomes a monotonic function and no more IMFs can be extracted.

**Figure 1. The EMD decomposition of a simple signal**

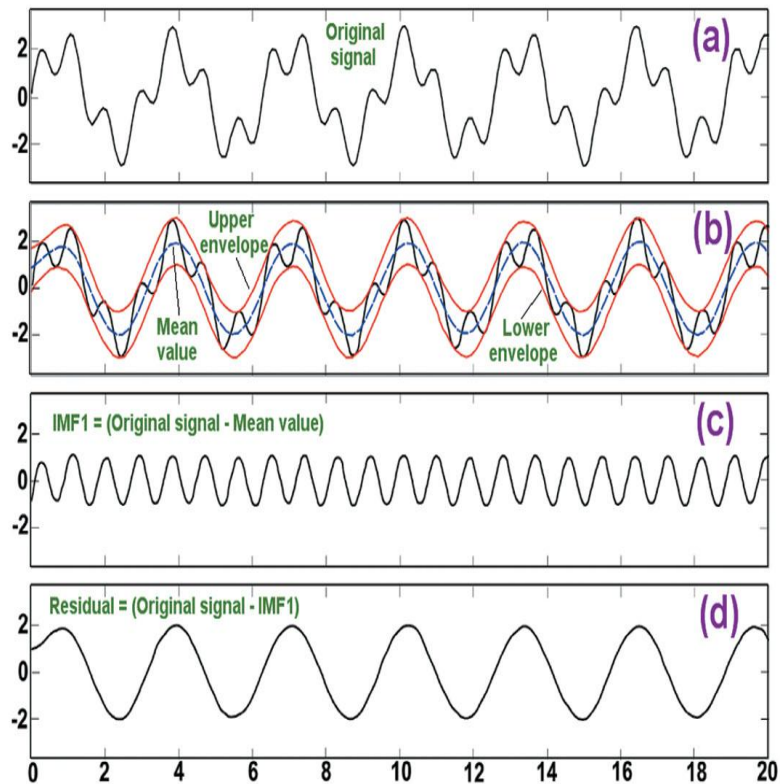

Notes: (a) the original signal, (b) the lower and upper envelopes (red) and their mean (blue), (c) the first IMF and (d) the first residual.

As effective as EMD proved to be, this new technique leaves some difficulties unresolved. One of the major shortcomings of the EMD is the mode mixing which is the result of signal intermittency. This may generate serious aliasing in the time-frequency distribution, and then the intrinsic modes might devoid of physical meaning. The Ensemble Empirical Mode Decomposition (EEMD), proposed by Wu and Huang (2009) as a truly noise-assisted data analysis, permits to avoid this problem. It defines the extracted IMFs as the means of an ensemble of trials. Each trial disentangles the signal and adding a finite amplitude white noise. The addition of noise could help data analysis in the EMD method and thus could minimize or overcome the mode-mixing problem. Figure 2 illustrates the transition from EMD to EEMD consisting of adding a White Gaussian Noise (AWGN), where the std is the noise amplitude in terms of standard deviations, and Ne is the ensemble number. To control for mode-mixing, the EEMD treats the EMD as basic function and repeats Ne times with distinct AWGNs.

**Figure 2. EMD versus EEMD**

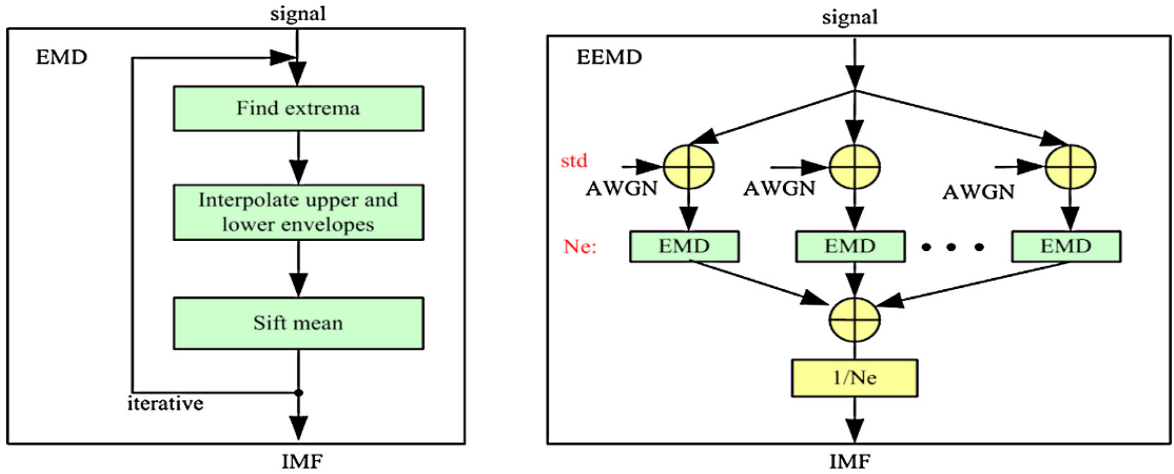

After the partition of the signal into a set of different frequency components by means of EEMD, a correlation analysis between Bitcoin price and U.S. stock price index and potential control variables is conducted. The model to be estimated is given by:

$$SPI_t = \omega + \alpha SPI_{t-1} + \beta BP_t + \gamma gold_t + \delta silver_t + \lambda WTI_t + \varepsilon_t \qquad (4)$$

Where SPI is the S&P 500 Stock Price Index; The S&P 500 Stock Price Index (SPI) covers the performance of 500 largest capitalization stocks, and is sourced from DataStream of Thomson Reuters; $SPI_{t-1}$: the lagged S&P 500 stock price that may reflect the influence of some potential explanatory variables not included here due to the unavailability of daily frequency data; BP is the Bitcoin price in US dollars collected from CoinDesk at www.coindesk.com/price.The CoinDeskBitcoin Price Index represents an average of Bitcoin prices across leading bitcoin exchanges; gold is the world gold price; silver is the world silver prices. Gold and silver are precious metals that have been largely viewed as safe haven over extreme stock market fluctuations (Baur and Lucey 2010). WTI is the West Texas Intermediate oil price. The WTI has been largely employed in the literature as the benchmark price for global oil markets. The WTI is among the most traded oil on the world markets, and therefore is significantly affected by macro-financial variables. WTI, gold and silver data come from quandl website; $\varepsilon_t$ is the error term.

We focus our analysis on the 2016 US presidential election outcome and address whether Bitcoin acts as a safe haven for U.S. stock price index in the aftermath of Trump's win. The final result of the election was disclosed on Tuesday 08 Nov 2016, which we subsequently view as the announcement day. So, this study uses daily data for the period that spans from 08 Nov 2016 to 15 Feb 2017 (99 observations).

3. Results

A multi-scale correlation analysis is conducted to assess whether Bitcoin serves a hedge or safe haven for U.S. stock index over the uncertainty surrounding the Trump's victory in U.S. presidential elections. The procedure consists of (1) decomposing the original time series into various intrinsic mode functions, and (2) examining the correlation between Bitcoin price and S&P500 index across the extracted components, and (3) addressing whether Bitcoin joins precious metals in safe-haven status, while controlling for gold and silver prices.

Based on EEMD, the Bitcoin price and S&P500 price index and the explanatory variables (gold and silver prices) are decomposed into five IMFs (Appendix A). The EEMD technique generates itself the modes depending to the data. All the derived IMFs are listed from high to low frequency components. We note that the frequencies and amplitudes of all the IMFs evolve over time. We discuss three scaling components: short-run (IMFs1-2: within one to two weeks), medium-run (IMFs 3-4: above two weeks and less than 12 weeks) and long-run (IMF5: above 12 weeks). Table 1 reports some measures which are given to assess IMFs: mean period of each IMF, correlation between each IMF and the original data series and the variance percentage of each IMF. The mean period corresponds to the value derived by dividing the total number of points by the number of peaks for each IMF. The Pearson and Kendall Rank coefficients are utilized to determine the correlation among IMFs and the original data. Because IMFs are intrinsically independent, it is possible to sum up the variances and employ the percentage of variance to measure the contribution of each IMF to the original data. The variables of interest (BP and SPI) are driven by the same components. In particular, both the U.S. stock price index and the Bitcoin price are explained by short-term hidden factors (IMF1 and IMF2).

**Table 1. The IMFs features**

|  | Mean period | Pearson correlation | Kendall correlation | variance as % of the sum of IMFs |
|---|---|---|---|---|
| BP | | | | |
| IMF1 | 3.62 | 0.262*** | 0.199** | 44.98% |
| IMF2 | 7.94 | 0.256* | 0.234** | 29.26% |
| IMF3 | 8.39 | 0.195*** | 0.167*** | 9.23% |
| IMF4 | 11.93 | 0.131** | 0.116** | 7.14% |
| IMF5 | 25.04 | 0.112** | 0.099* | 6.68% |
| SPI | | | | |
| IMF1 | 4.32 | 0.183* | 0.175** | 36.98% |
| IMF2 | 7.69 | 0.277** | 0.259** | 41.12% |
| IMF3 | 8.16 | 0.113* | 0.089** | 1.18% |
| IMF4 | 13.18 | 0.218*** | 0.204*** | 3.46% |
| IMF5 | 38.25 | 0.096** | 0.075*** | 17.27% |

Notes: *, **, ***: Correlations are significant at the levels of 0.01, 0.05 and 0.1.

Figure 3 confirms the previous results showing that the high frequency components are the major contributors of BP and SPI.

**Figure 3. The hidden characteristics of Bitcoin price and U.S. stock price index**

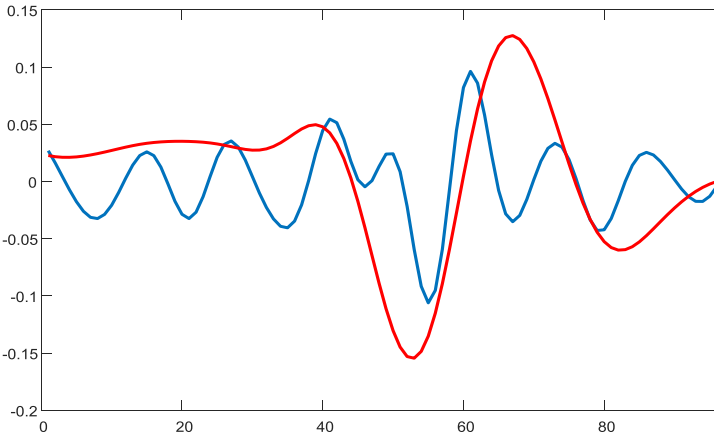

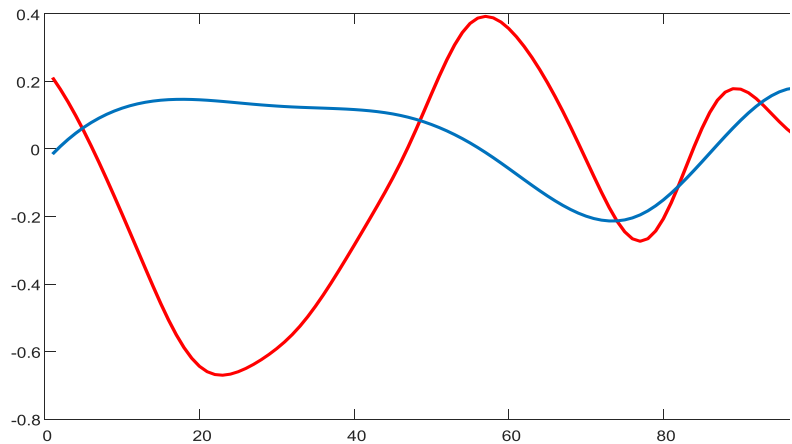

Notes: the red line represents the high frequency component; the blue line represents the low frequency component.

To test if Bitcoin joins precious metals in safe haven properties in the aftermath of 2016 U.S presidential elections outcomes, we have estimated the multi-scale correlation between Bitcoin price and S&P500 stock price index, even if we control for gold and silver prices. We define a safe haven depending on the sign of the coefficients associated to the gold price. If the coefficient is negative or insignificant (i.e., uncorrelated), then Bitcoin, gold or silver may be served as a hedge or a safe haven. We include also WTI as a potential fundamental of stock price valuation. Table 2 reports the OLS-based EEMD regression' results.

During the post-2016 U.S. election period, Bitcoin price and S&P 500 index are likely to be negatively correlated in the short and the long term, but uncorrelated in the medium-run. This result underscores a new confidence in Bitcoin as a safe haven. As is with every other political issue regarding Trump's administration, it remains unclear what to expect. But there appears to be a quite general consensus in the Bitcoin community that whatever Trump's policies turn out to be, Bitcoin will benefit largely. This may explain the supported Bitcoin safe haven property. The reason that is making Bitcoiners hopeful about Trump's triumph is the inclusion of Bitcoin supporters like Peter Thiel, Balaji Srinivasan and Mick Mulvaney in his team. Peter Thiel is a technology entrepreneur and investor; he is the co-founder of PayPal, a Bitcoin enthusiast and has invested into multiple Bitcoin companies. Balaji Srinivasan is one of the best-funded Bitcoin startups so far. He is the co-founder and CEO of 21; the latter has developed a full stack set of technologies for practical Bitcoin micropayments. Also, Mick Mulvaney, the designated Director of the Office of Management and Budget under Trump's presidency, is viewed as one of the most representatives of the crypto community since he is more outspoken about Blockchain technology and Bitcoin. Having Bitcoin believers in the Trump's team is a win for the Bitcoin community in the whole. More interestingly, Donald Trump's election victory has sent U.S markets on a tumultuous ride. Markets are reacting as investors find out how heavy are the president-elect's statements on trade, fiscal policy and regulation. And as experienced after Trump's win, the uncertainty can encourage people to hoard assets such as Bitcoin and precious metals, both perceived as a hedge against uncertainty. But considering the sizable market volatility and the changes in traders' attitudes, it's natural to wonder whether Bitcoin can be viewed as a purely safe haven, which may explain the decreasing correlation between Bitcoin price and U.S stock price over time.

Unlike Bitcoin, the precious metals (gold and with less extent silver) exert a non-negative influence on the U.S. stock price index in the short and medium-run (IMF1, IMF2 and IMF3) and negative effect in the long term (IMF4 and IMF5). This means that the property of gold as well as silver is time-varying. In particular, they act as weak safe haven for S&P500 price index in short and intermediate time scales, and as a hedge in the longer time horizons. Typically, when the economy witnessed an evolving volatility that may impede stocks' valuation, investors may shift their funds from stocks and invest them in the gold and silver markets until the economy rebounds. In this context, precious metals could act as a stabilizer control in investment portfolios, and play as safe haven during turbulent times (Baur and Lucey 2010). Also, traders tend to go into more liquid assets when market turbulence emerges and uncertainty increases. As an extremely liquid asset even in periods of market turmoil, gold can be served as a hedge.

Furthermore, there are swings from positive to negative correlation between WTI and S&P 500 over time, reflecting the volatile and the speculative behavior of World oil market. This finding seems against the alternative of adding crude oil to serve as a commodity diversifier in uncertain context.

**Table 2. Multi-scale estimates of the relationship between Bitcoin price and U.S. stock price**

|        | IMF1      | IMF2      | IMF3      | IMF4      | IMF5      |
|--------|-----------|-----------|-----------|-----------|-----------|
|        | Dependent variable: SPI ||||| 
| C      | -0.5408*  | 0.9785**  | 0.663966  | 0.796386* | 1.108502  |
|        | (0.0439)  | (0.0041)  | (0.1700)  | (0.0304)  | (0.1989)  |
| SPI(-1)| -0.27057* | -1.059**  | -0.32362* | 0.255759* | 0.235810  |
|        | (0.0426)  | (0.0071)  | (0.0345)  | (0.0855)  | (0.6221)  |
| BP     | -0.141**  | -0.129**  | 0.543518  | -0.14424  | -0.1102*  |
|        | (0.0007)  | (0.0095)  | (0.4610)  | (0.2131)  | (0.3202)  |
| gold   | 0.05799*  | 0.0417*   | 0.0278*   | -0.0936** | -0.0931*  |
|        | (0.0153)  | (0.0173)  | (0.0306)  | (0.0016)  | (0.0460)  |
| silver | 0.0446**  | 0.0221*   | 0.012213* | -0.0036** | -0.0161*  |
|        | (0.0050)  | (0.0235)  | (0.0164)  | (0.0036)  | (0.0761)  |
| WTI    | -0.09861* | 0.1158*   | 0.09687*  | -0.1410** | 0.0984*   |
|        | (0.0865)  | (0.0137)  | (0.0672)  | (0.0059)  | (0.0126)  |
| R2     | 0.85      | 0.83      | 0.84      | 0.86      | 0.85      |

Notes: (.): p-values. ***, ** and * in the table denote statistical significant coefficients at 1 per cent, 5 per cent and 10 per cent level, respectively.

## 4. Robustness

There exist different ways to check the robustness of our results. We carried out a series of robustness checks. First, we re-examine the correlation while replacing the nominal stock price by the real stock prices[4]. While stock prices are heavily determined by financial variables (Valcarcel 2012), the prominence of macroeconomic variables cannot be ruled out (Goyal and Welch 2008; Rapach and Zhou 2013). Inflation is considered as one of the most potential macroeconomic variables believed to be related to stock prices. Even though inflationary shocks may have weak long-run impact on stock returns owing to monetary non-neutrality, it was largely documented that stock prices can be influenced by inflation in the short-run (Valcarcel 2012; Bjørnland and Jacobsen 2013). This underscores the importance to replace the nominal stock price by the real stock prices while adjusting the stock market pricing for inflation. This task may allow us to address whether the U.S monetary policy can stimulate the performance of stock market over the great uncertainty surrounding the U.S presidential elections. The results are reported in Table 3. Using another U.S. stock price proxy, we confirm that:

(i) The hedge and safe-haven properties of Bitcoin, gold and silver vary depending to different time-horizons.

(ii) Bitcoin poorly acts as a safe haven in the short and the long-run.

(iii) The gold and with less extent silver appear as interesting assets to hedge against unexpected events and uncertainty in the short-term. But they lost their hedging and safe haven properties in the long-run.

(iv) The viability of oil as diversifier for a portfolio that is formed by U.S stocks remains unproven.

---

[4] Since inflation rate is unavailable in daily frequency, we use Data Frequency Conversion from low (i.e., inflation monthly data) to high (i.e., inflation daily data) frequency via Eviews 9. For more details about the conversion procedure, you can refer to this link: http://www.eviews.com/Learning/freqconv.html. Monthly inflation data are collected from EcontatsTM.

**Table 3. Multi-scale estimates of the relationship between Bitcoin price and real U.S. stock price**

|        | IMF1       | IMF2       | IMF3       | IMF4       | IMF5       |
|--------|------------|------------|------------|------------|------------|
|        | Dependent variable: real SPI ||||||
| C      | 0.565629   | 0.891338*  | -0.17745*  | -0.0981**  | -0.50874*  |
|        | (0.9331)   | (0.0315)   | (0.0597)   | (0.0079)   | (0.0719    |
| SPI(-1)| -0.088349  | 0.054829   | -0.01213*  | 0.1181*    | 0.125603   |
|        | (0.8403)   | (0.2674)   | (0.0538)   | (0.0556)   | (0.7987    |
| BP     | -0.14348** | -0.15141*  | -0.08592   | -0.1433*   | -0.12482*  |
|        | (0.0095)   | (0.0636)   | (0.3490)   | (0.0116)   | (0.0170)   |
| gold   | 0.094893*  | 0.05097**  | -0.108265  | -0.135**   | 0.092123   |
|        | (0.0216)   | (0.0042)   | (0.3479)   | (0.0058)   | (0.1632)   |
| silver | 0.049722*  | 0.028905*  | 0.02015**  | -0.0863*   | -0.0663**  |
|        | (0.0343)   | (0.0474)   | (0.0091)   | (0.0486)   | (0.0054)   |
| WTI    | -0.10139*  | 0.0529**   | -0.1235**  | -0.1164*   | 0.02374**  |
|        | (0.0835)   | (0.0016)   | (0.0021)   | (0.0111)   | (0.0069)   |
| R2     | 0.90       | 0.84       | 0.89       | 0.90       | 0.89       |

Notes: (.): p-values.  ***, ** and * in the table denote statistical significant coefficients at 1 per cent, 5 per cent and 10 per cent level, respectively.

Second, we replace the spot prices of gold and silver by their future prices. It must be stressed at this stage that the futures prices of precious metals are cleared via a centralized exchange with standardized contracts and complete fiduciary transparency. With heavily regulated markets, investors have access nearly 24hrs by day. On the contrary, the spot prices of the metals under study are completely unregulated. In particular, spot gold and silver dealers undertake their own prices and policies. In this way, these prices do not often reflect the world gold and prices.

The spot and futures prices are closely related via arbitrage strategies. For financial assets, if all arbitrage is possible, the forward price $F_t^{t+1}$ in $t$ for $t+1$, is equal to the spot price $S_t$ at date t and the interest cost foregone over the period ($r_t S_t$).

$$F_t^{t+1} = S_t + r_t S_t \tag{5}$$

Since the storage costs must be accounted for (Fama and French 1988), we followed Coudert and Raymond (2010) by including the cost of physical storage $w_t$ and by deducting then the convenience yield $c_t^5$ to the forward price yet defined in Equation (5).

$$F_t^{t+1} = S_t + r_t S_t + w_t - c_t \tag{6}$$

Using two different alternative prices of gold and silver, we robustly show that:

(i) Bitcoin may serve as a weak safe haven in the short-run and the long-run (negative correlation), and as a hedge in the medium term (uncorrelated with U.S. stock price).

(ii) Precious metals lost their safe haven properties as their correlation with U.S. stock price index are declining from one frequency to another.

(iii) Including crude oil in a portfolio formed by U.S stock price to serve as a commodity diversifier in turbulent times remains unsupported.

---

[5] The convenience yield represents the gain obtained through storing the commodity, and that stems from the uncertainty surrounding the future production. For the future prices of gold and silver, the data are available in Datastream database.

**Table 4. Multi-scale estimates of the relationship between Bitcoin price and U.S. stock price by replacing the spot prices of gold and silver by the forward prices**

|        | IMF1       | IMF2      | IMF3      | IMF4       | IMF5       |
|--------|------------|-----------|-----------|------------|------------|
|        | Dependent variable: SPI ||||| 
| C      | -0.40923*  | 0.521058  | -0.4009*  | -0.654***  | 0.21386**  |
|        | (0.0352)   | (0.3894)  | (0.0305   | (0.0000)   | (0.0037)   |
| SPI(-1)| -0.204381  | 0.045861  | -0.21213* | 0.19864*   | 0.28438**  |
|        | (0.9210)   | (0.1527)  | (0.0326)  | (0.0425)   | (0.0045)   |
| BP     | -0.148***  | -0.12351* | 0.1939    | -0.096***  | -0.1114**  |
|        | (0.0000)   | (0.0950)  | (0.2385)  | (0.0000)   | (0.0095)   |
| gold   | 0.0746***  | 0.13451   | 0.45216   | -0.0613**  | -0.0341*** |
|        | (0.0002)   | (0.1843)  | (0.5133)  | (0.0045)   | (0.0001)   |
| silver | 0.0446**   | 0.0221*   | 0.01218   | -0.0356*   | -0.0161*   |
|        | (0.0050)   | (0.0235)  | (0.3164)  | (0.0145)   | (0.0761)   |
| WTI    | -0.1056**  | 0.08627*  | -0.1478*  | -0.119***  | 0.09653*** |
|        | (0.0031)   | (0.0290)  | (0.0997)  | (0.0000)   | (0.0001)   |
| R2     | 0.85       | 0.83      | 0.84      | 0.86       | 0.85       |

Notes: (.): p-values. ***, ** and * in the table denote statistical significant coefficients at 1 per cent, 5 per cent and 10 per cent level, respectively.

## 5. Discussion and conclusion

Donald Trump is one of the most controversial figures of the current times. So much so, that his victory in the U.S. presidential elections had led to a great uncertainty in the world economy. In the aftermath of Trump's win, a particular interest was devoted to digital assets (in particular, Bitcoin), strengthening its status as the modern safe haven. As safe havens and hedge rely on co-movements between assets in times of market turmoil, this paper seeks to examine the time-varying relationship between Bitcoin price and U.S. stock price index during the post-U.S. election period. We then compare Bitcoin hedging and safe haven properties to the traditional assets (in particular, gold, silver).

An effective analysis of such complex issue needs the exclusion of the parametric approaches that use predetermined patterns. The core focus is to suggest a newly and adequate econometric tool for reading data and permitting the investigation with respect to different frequencies, namely Ensemble Empirical Mode Decomposition. Indeed, the price of several assets (Bitcoin, U.S. stocks, gold, silver and oil) data were decomposed into several intrinsic mode functions, even if we control for aliasing in the time-frequency distribution.

Our results indicate that the hedge and safe-haven properties of Bitcoin, gold and silver for U.S. stock market is not constant over time. Bitcoin acts as a weak safe haven in the short-run, and as a hedge in the medium- and the long-run. We show also that gold and silver lost their hedge and safe haven properties as the co-movement among precious metals and U.S. stock price index decreases over time. Investors and traders are generally interested in hedges that mitigate the volatility of their portfolio, but also they are likely interested in buying some sort of insurance against extreme tail events such as the uncertainty surrounding the unanticipated U.S. election outcome. Bitcoin has several properties that make it a very interesting asset in both cases. Currently, the loss of faith in the stability of banking system and future economic security worsened, and market uncertainty heightened across the globe. But Bitcoin which lives outside the confines of a single country's politics has profited from the current ongoing volatility. While these properties may justify that Bitcoin serves as a hedge for U.S. stock price index in turbulent times, they explain also why Bitcoin cannot be considered as a strong safe haven. When market turmoil arises and uncertainty rises, investors are known to sell "risky" assets and buy "safer" assets, also known as "flights to safety" (Baele et al. 2015). And Bitcoin does not hold this property. Being a crypto-currency, Bitcoin is highly sensitive to cyber-attacks which may destabilize its whole system (Barber et al. 2012).

Further, Bitcoin and precious metals do not act in similar way for U.S. stock market. From a legal perspective, Bitcoin does not appear to share the features of traditional safe-haven investments. Specifically, the validity of Bitcoin as a hedge and a safe haven may encounter a number of obstacles. Although Bitcoin is a liquid asset even in times of market upheaval, it is a high-risk, volatile and speculative investment. However, as extremely liquid assets even in turbulent times, gold and silver can be characterized as safer assets.

**References**

Baele, L., Bekaert, G., Inghelbrecht, K. and Wei, M., (2015). 'Flights to Safety'. Working Paper.

Baur, D. and Lucey, B. (2010).Is gold a Hedge or Safe Haven? An Analysis of Stocks Bonds and Gold. The Financial Review 45: 217–229.

Baur, D.G. and McDermott, T.K., (2010). Is gold a safe haven? International evidence. Journal of Banking and Finance 34, 1886-1898.

Baur, D. G., Lee, A. D. and Hong, K. (2015). Bitcoin: Currency or Investment?.Available at SSRN 2561183.

Bjørnland, H.C., and Leitemo K., 2009. Identifying the Interdependence between US Monetary Policy and the Stock Market. Journal of Monetary Economics, 56, 275-282.

Bouoiyour, J. and Selmi, R. (2015). What Does Bitcoin Look Like? Annals of Economics and Finance 16(2): 449-492.

Bouoiyour, J., Selmi, R. and Tiwari, A. K. (2015). Is Bitcoin business income or speculative foolery? New ideas through an improved frequency domain analysis.Annals of Financial Economics, 10(01), 1550002.

Bouoiyour, J., Selmi, R., Tiwari, A-K. and Olayeni, O-R., (2016). What drives Bitcoin price? Economics Bulletin 36(2): 843-850.

Bouoiyour, J., and Selmi, R., (2016). The winners from Brexit — on the safe haven properties of Bitcoin. Working paper, CATT, University of Pau.

Bouri, E., Molnár, P., Azzi, G., Roubaud, D. and Hagfors, L. I. (2015). On the hedge and safe haven properties of Bitcoin: Is it really more than a diversifier? Finance Research Letters, doi:10.1016/j.frl.2016.09.025.

Brandvold, M., Molnár, P., Vagstad, K., and Valstad, O. C. A. (2015). Price discovery on Bitcoin exchanges. Journal of International Financial Markets, Institutions and Money 36: 18–35.

Buchholz, M, Delaney, J, Warren, J, and Parker, J, (2012). Bits and Bets, Information, Price Volatility, and Demand for Bitcoin. Economics 312, http://www.bitcointrading.com/pdf/bitsandbets.pdf

Ciaian, P. Rajcaniova, M. and Kancs, D.A. (2016). The economics of BitCoin price formation. Applied Economics 48(19): 1799-1815.

Coudert, V., and Raymond, H., (2010). Gold and financial assets: are there any safe havens in bear markets? CEPII working paper n° 2010-13, July.

Dyhrberg, A.H. (2016). Hedging capabilities of Bitcoin. Is it the virtual gold? Finance Research Letters 16: 139-144.

Fama, E.F. and French, K.R. (1988). Business cycles and the behavior of metal prices. Journal of Finance 43: 1075–1088.

Garcia, D., Tessone, C., Mavrodiev, P. and Perony, N. (2014). The digital traces of bubbles: feedback cycles between socio-economic signals in the Bitcoin economy. Journal of the Royal Society Interface 11(99): 1-8.

Goyal, A. and Welch, I., 2008. A Comprehensive Look at the Empirical Performance of Equity Premium Prediction. Review of Financial Studies, 21(4) 1455-1508.

Huang, N. E., Z. Shen, and S. R. Long, M. C. Wu, E. H. Shih, Q. Zheng, C. C. Tung, and H. H. Liu, (1998), The empirical mode decomposition method and the Hilbert spectrum for non-stationary time series analysis. Proc. Roy. Soc. London, 454A: 903-995.

Kristoufek, L., (2013). Bitcoin meets Google Trends and Wikipedia: Quantifying the relationship between phenomena of the Internet era. Scientific Reports 3 (3415): 1-7.

Polasik, M., Piotrowska, A., Wisniewski, T. P., Kotkowski, R. and Lightfoot, G. (2014). Price Fluctuations and the Use of Bitcoin: An Empirical Inquiry. Available at SSRN 2516754.

Popper, N. (2015). Digital gold: The untold story of Bitcoin. London: Penguin.

Rapach, D., and Zhou, G., (2013). Forecasting Stock Returns. Handbook of Economic Forecasting, Volume 2A, Graham Elliott and Allan Timmermann (Eds.) Amsterdam: Elsevier, 328–383.

Valcarcel, V. J., (2012). The dynamic adjustments of stock prices to inflation disturbances. Journal of Economics and Business, 64, 117– 144.

Wu, Z. and N. E. Huang, (2003). A study of the characteristics of white noise using the empirical mode decomposition method. Proc. Roy. Soc. London, 460A, pp. 1597-1611.

Wu, Z.H., and N.E. Huang, N.E. (2009). Ensemble empirical mode decomposition: A noise-assisted data analysis method. Advances in Adaptive Data Analysis 1(1): pp. 1-41.

Yermack, D. (2013). Is Bitcoin a real currency? An economic appraisal. NBER working paper n° w19747.

# Appendix

## Bitcoin price and S&P 500 index: IMFs derived from EEMD

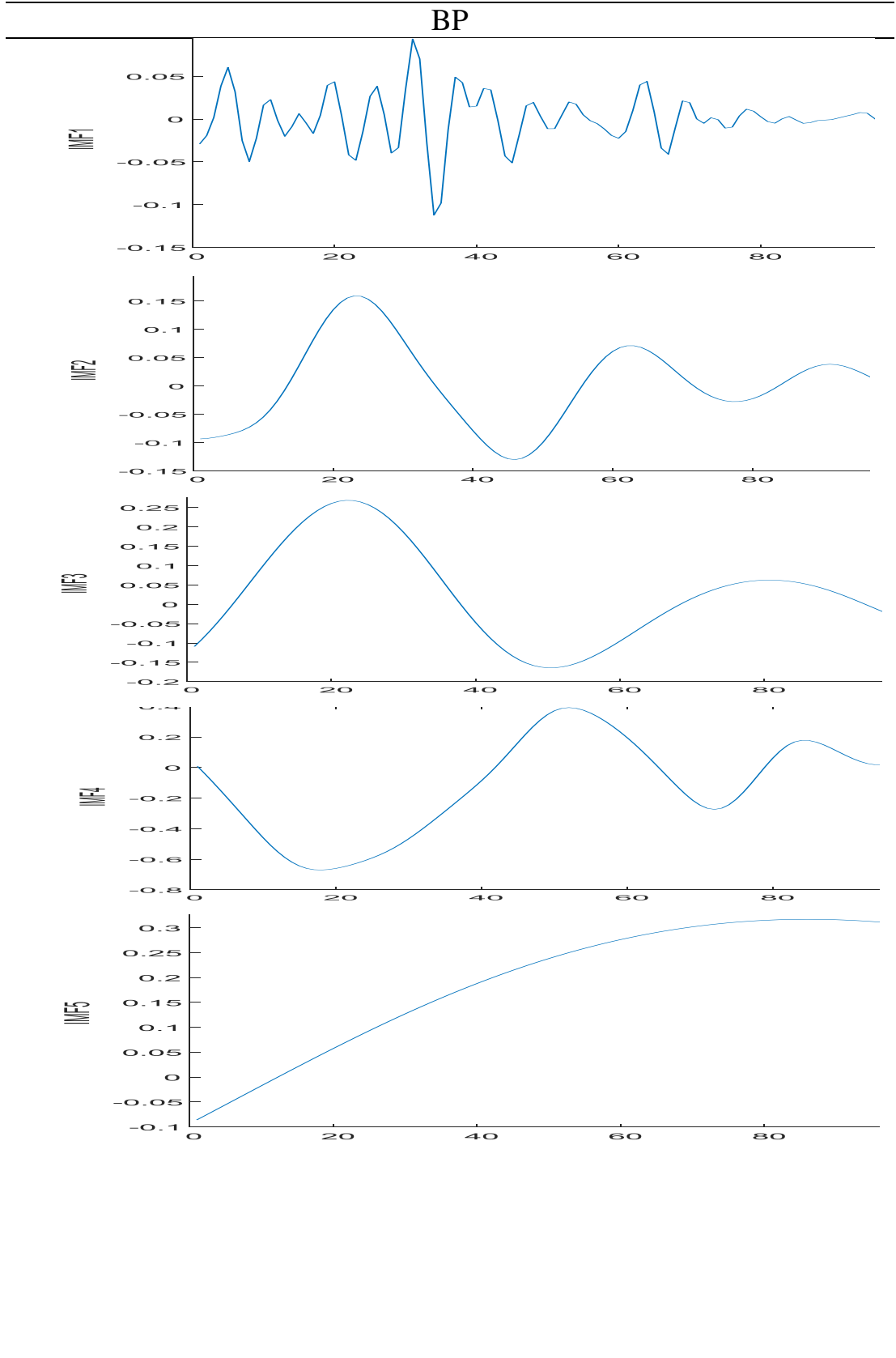

BP

SPI

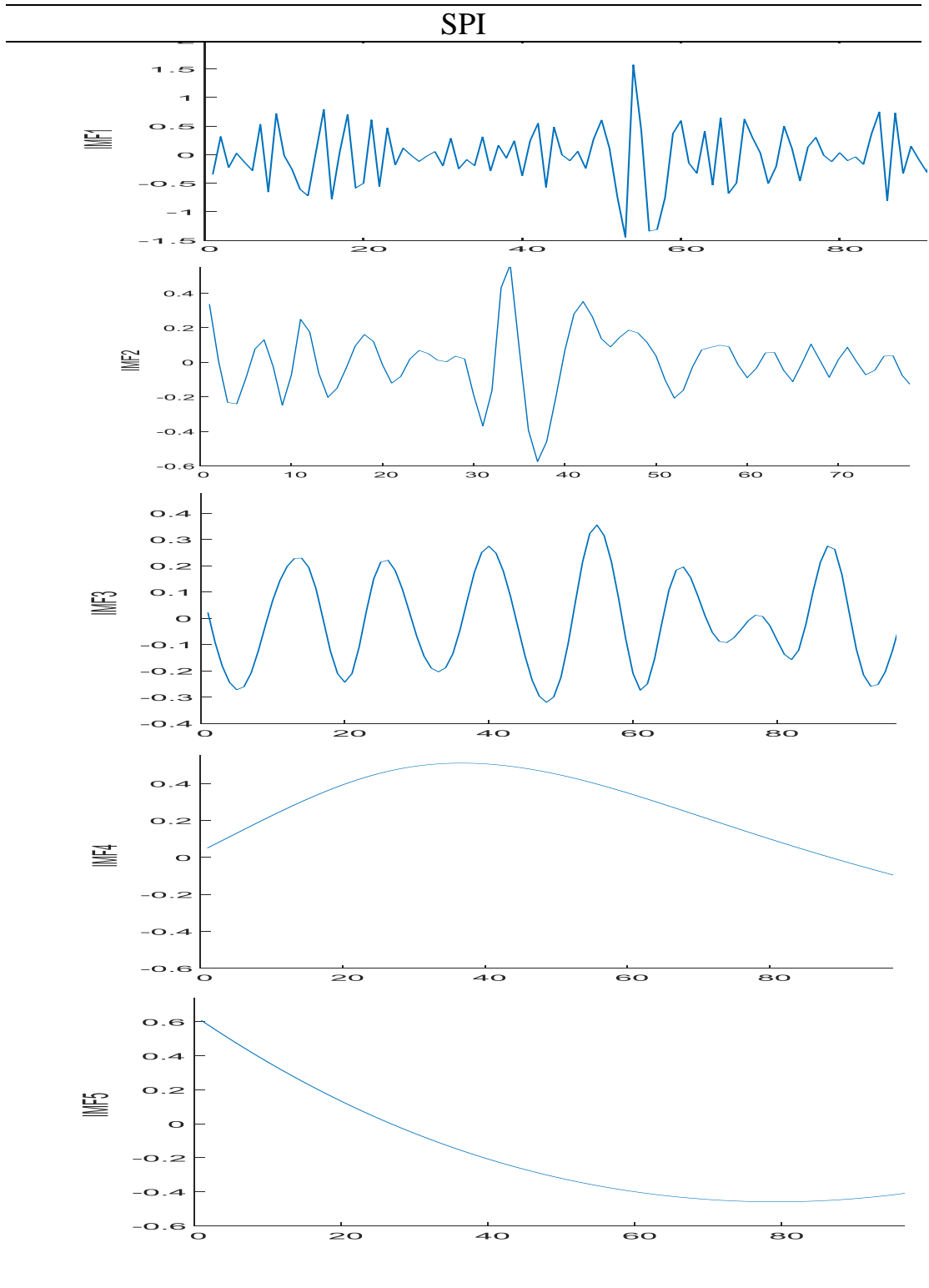